\documentclass{article}
\usepackage{template}

\usepackage{times}
\usepackage{soul}
\usepackage{url}
\usepackage[hidelinks]{hyperref}
\usepackage[utf8]{inputenc}
\usepackage[small]{caption}
\usepackage{graphicx}
\usepackage{amsmath}
\usepackage{amssymb}
\usepackage{amsfonts}
\usepackage{amsthm}
\usepackage{booktabs}
\usepackage{algorithm}
\usepackage{algorithmic}
\usepackage{bm}
\usepackage{bxtexlogo}
\usepackage{float}
\usepackage{graphicx}
\usepackage{mathtools}
\usepackage{multirow}
\usepackage[switch]{lineno}

\usepackage{tabularx}
\usepackage{xcolor}

\newcommand{\takehiro}[1]{\textcolor{black}{#1}}
\newcommand{\hashimoto}[1]{\textcolor{black}{#1}}

\title{CoFinDiff: Controllable Financial Diffusion Model for Time Series Generation}

\author{
Yuki Tanaka$^1$
\and
Ryuji Hashimoto$^1$
\and
Takehiro Takayanagi$^1$
\and
Zhe Piao$^2$
\and \\
Yuri Murayama$^1$
\And
Kiyoshi Izumi$^1$
\affiliations
$^1$The University of Tokyo\\
$^2$Nomura Holdings, Inc.\\ 
}

\begin{document}

\maketitle

\begin{abstract}
The generation of synthetic financial data is a critical technology in the financial domain, addressing challenges posed by limited data availability. Traditionally, statistical models have been employed to generate synthetic data. However, these models fail to capture the stylized facts commonly observed in financial data, limiting their practical applicability. Recently, machine learning models have been introduced to address the limitations of statistical models; however, controlling synthetic data generation remains challenging. 
We propose CoFinDiff (\textbf{Co}ntrollable \textbf{Fin}ancial \textbf{Diff}usion model), a synthetic financial data generation model based on conditional diffusion models that accept conditions about the synthetic time series. 
By incorporating conditions derived from price data into the conditional diffusion model via cross-attention, CoFinDiff learns the relationships between the conditions and the data, generating synthetic data that align with arbitrary conditions. Experimental results demonstrate that: (i) synthetic data generated by CoFinDiff capture stylized facts; (ii) the generated data accurately meet specified conditions for trends and volatility; (iii) the diversity of the generated data surpasses that of the baseline models; and (iv) models trained on CoFinDiff-generated data achieve improved performance in deep hedging task.

\end{abstract}

\section{Introduction}

\begin{figure*}[tbph]
  \centering
  \includegraphics[width=.9\linewidth]{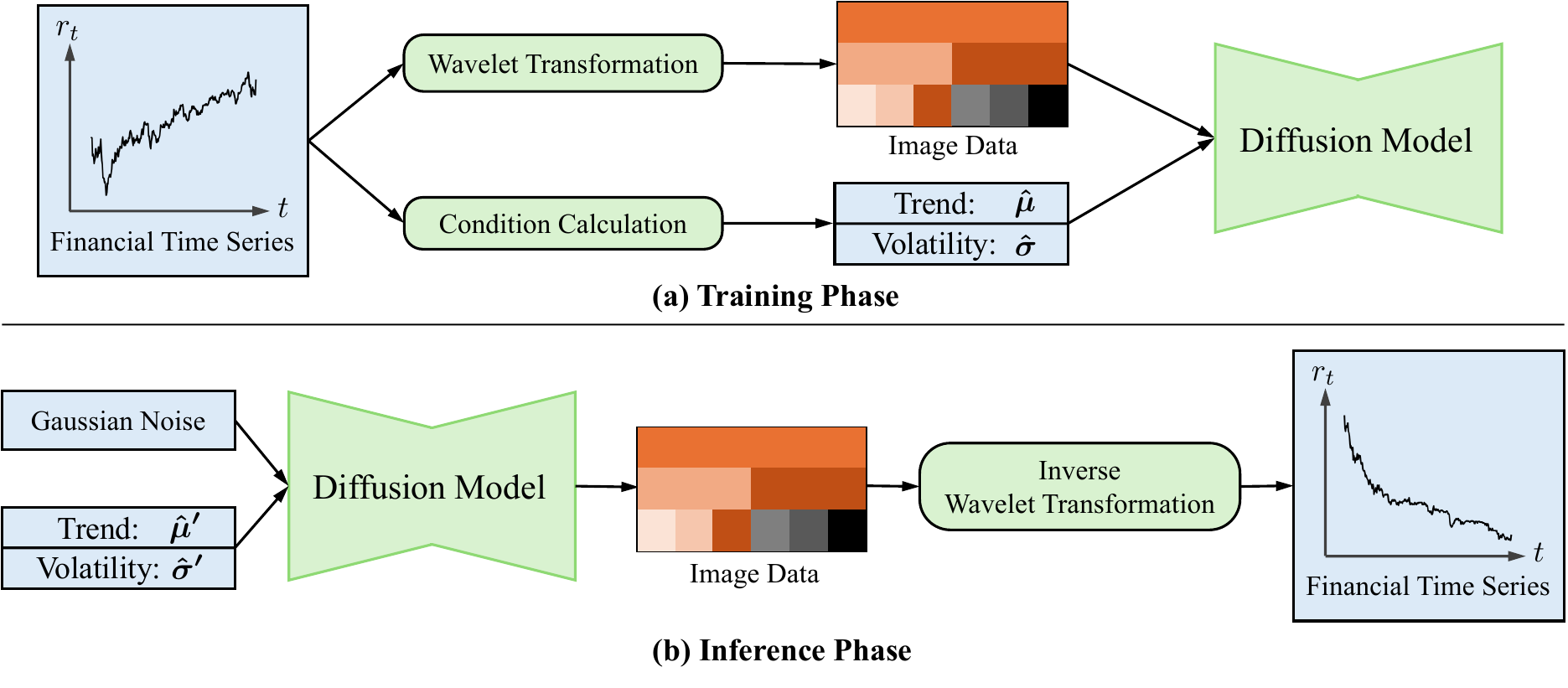}
  \caption{\hashimoto{CoFinDiff overview.
  (a) In the training phase, the diffusion model learns to reproduce input logarithmic return series (Sec. \ref{Sec:training_phase}). The input data is converted into images using the Haar wavelet transformation, and the trend and realized volatility are computed to serve as conditions fed into the cross-attention mechanism of the diffusion model (Sec. \ref{Sec:input_data_processing}). (b) In the inference phase, CoFinDiff accepts arbitrary condition by a modeler. The model outputs an image, which is then converted back to a logarithmic return series that adheres to the specified conditions using reversed wavelet transformation (Sec. \ref{Sec:inference_phase}).
  }
}
  \label{Fig:overview}
\end{figure*}


\takehiro{Synthetic financial data generation is a critical technology with diverse applications in financial markets. For instance, automated trading systems, a major application of machine learning (ML) in the financial domain, require diverse data sets to be well-prepared for a variety of future scenarios. However, financial data such as price series accumulate gradually over time, which limits the availability of data for these scenarios~\cite{synthetic_data_overview}. As a result, to leverage ML-based automated trading technology, synthetic financial data generation has gained considerable research attention.}


\takehiro{Traditionally, synthetic financial data generation has relied on statistical models like Wiener processes, ARCH, and GARCH~\cite{engle1982autoregressive,bollerslev1986generalized}. However, these models lack the expressive power to capture all stylized facts\footnote{Stylized facts refer to the commonly observed features of financial data across various asset classes, such as fat tails and volatility clustering.}~\cite{malmsten2010stylized}, causing synthetic data to diverge from reality.}


To overcome these limitations, recent research has turned to deep generative models. 
The high expressiveness of deep generative models enables the generation of data that satisfy stylized facts, as demonstrated by examples such as variational autoencoders (VAEs)~\cite{vae}, generative adversarial networks (GANs)~\cite{gan}, and diffusion models~\cite{ho2020denoising}.

\takehiro{
While successful in realistic data generation, previous research has largely overlooked the controllability of synthetic data generation in finance, despite its critical role in financial applications. Financial markets exhibit fat-tail distributions~\cite{stylized_facts}, which may trigger extreme events such as flash crashes.\footnote{A flash crash is defined as a sudden, severe drop in asset prices that recovers rapidly~\cite{kirilenko2017flash}.} These rare events carry significant risks, yet their scarcity means that relevant data are extremely limited. Therefore, addressing data scarcity in finance necessitates the conditional generation of specific events, including extreme events. Recent studies have explored conditional GANs~\cite{mirza2014conditionalgenerativeadversarialnets,gan_time_series,li2020generating,wiese2020quant} to generate synthetic financial data conditioned on historical data. While effective at capturing historical trends, this approach restricts controllability because it does not allow for the direct specification of unique market conditions. Thus, no studies have been conducted on conditional generation methods that directly specify data conditions (e.g., high volatility accompanied by a very sharp downtrend).}

Aiming to promote the use of ML technologies in finance by generating synthetic \takehiro{financial data that capture stylized facts and offer controllability}, this study proposes CoFinDiff, a conditional synthetic financial data generation model using conditional diffusion models. CoFinDiff transforms stock price series into images using the Haar wavelet~\cite{wavelet} and employs conditional diffusion models for data learning and generation. By incorporating trend and realized volatility, calculated from the stock price series as conditions during the learning and generation processes, CoFinDiff can generate stock price series with arbitrary trends and realized volatility. Figure~\ref{Fig:overview} illustrates the overview of our proposed model.

\hashimoto{In our experiments, we assess the following four research questions.}\\ \indent
\textbf{RQ 1}: Do the synthetic data generated by CoFinDiff satisfy the stylized facts? \\ \indent
\textbf{RQ 2}: Can CoFinDiff generate synthetic data following conditions input into a model? \\ \indent
\textbf{RQ 3}: Are the synthetic data generated by CoFinDiff well diversified even under the trend and volatility constraints? \\ \indent
\textbf{RQ 4}: Are the synthetic data generated by CoFinDiff effective for downstream task?\\
\hashimoto{The experimental results suggested that CoFinDiff (i) generates data that aligns with stylized facts including fat tail and volatility clustering; (ii) accurately adheres to specified conditions; (iii) \takehiro{generates} more diverse synthetic data under fixed conditions than baselines including 
\takehiro{conditional} GANs; and (iv) improves the performance of deep hedging models by providing abundant synthetic data, especially for scenarios that are rarely included in real data.}







\section{Related Works}

Synthetic financial data have emerged as a promising tool for accelerating ML applications in finance, particularly in automated trading~\cite{synthetic_data_overview}. Various synthetic data generation methods have been developed to mitigate the limitations of real-world data availability.
\hashimoto{In particular,} data-driven techniques based on deep learning have gained traction. For instance, VAEs~\cite{vae} and GANs~\cite{gan} have been applied to generate synthetic financial time series, demonstrating improved performance in price prediction models~\cite{vae_gan}. GANs have been applied to generating correlated stock prices~\cite{gan_multiple}, synthesizing price series for deep hedging~\cite{deephedging_gan}, modeling continuous time series~\cite{gan_time_series}, and augmenting training data for reinforcement learning-based trading models~\cite{gan_to_rl}. Recently, diffusion models have shown promise in generating synthetic financial tabular data~\cite{findiff}. 
Although both statistical and deep learning-based models effectively leverage real data, no generation method exists for synthetic financial data generation that captures stylized facts and enables controllability.

The primary application of synthetic financial data is deep hedging, a type of automated trading system.
Deep hedging tasks, which are one of the primary applications in automated trading systems, represent a major application domain for synthetic financial data.
Deep hedging tasks are major application domains of synthetic financial data.
In \cite{deep_hedging}, synthetic data generated via the GJR-GARCH model was employed to train a model for constructing a hedging portfolio. 
\cite{mikkila2023empirical} argues that training hedge strategies using artificial data based on specific statistical models is ineffective, and that training with real data yields more effective results \cite{mikkila2023empirical}. Additionally, \cite{hirano2023adversarial} employed deep generative models to learn unnecessary hedge strategies for asset price processes, achieving promising results though hedge performance did not necessarily improve.

\section{CoFinDiff}

{
\begin{table*}[t]
\centering
\caption{Computed statistics representing stylized facts for both real and synthetic data. Terms in parentheses correspond to the standard deviation of each statistic. Real data comprise individual stock data, while synthetic data are generated using GBM, GARCH model, GANs, and the proposed method. "Hill" refers to the Hill Index and "Acorr" denotes autocorrelation. 
Check marks indicate that the corresponding stylized fact is satisfied.}
\small
\begin{tabular}{@{\extracolsep{\fill}}r|lllll|l}
\toprule
          & Real data (7203) & GBM & GARCH (1, 1) & vanilla GAN & Wasserstein GAN & \textbf{CoFinDiff} \\
\midrule
Kurtosis  &$6.97~(\pm10.60)$&$0.00~(\pm0.13)$&$0.18~(\pm0.46)$& $5.39~(\pm4.97)$           & $4.27~(\pm4.91)$          &$5.16~(\pm5.13)$\\
Hill      &$3.07$          &$5.97$          &$5.42$          & $3.06$           & $3.40$          &$2.98$\\
Acorr (1) &$0.19~(\pm0.10)$&$0.00~(\pm0.06)$&$0.02~(\pm0.07)$& $0.24~(\pm0.10)$          & $0.21~(\pm0.10)$         &$0.20~(\pm0.10)$\\
Acorr (5) &$0.13~(\pm0.08)$&$0.00~(\pm0.06)$&$0.06~(\pm0.07)$& $0.17~(\pm0.09)$           & $0.15~(\pm0.09)$          &$0.16~(\pm0.09)$\\
Acorr (10)&$0.10~(\pm0.08)$&$0.00~(\pm0.06)$&$0.03~(\pm0.06)$& $0.15~(\pm0.08)$           & $0.10~(\pm0.08)$          &$0.12~(\pm0.08)$\\
Acorr (20)&$0.07~(\pm0.07)$&$0.00~(\pm0.06)$&$0.00~(\pm0.06)$& $0.08~(\pm0.06)$           & $0.06~(\pm0.07)$          &$0.08~(\pm0.07)$\\
Acorr (30)&$0.06~(\pm0.07)$&$0.00~(\pm0.06)$&$0.00~(\pm0.06)$& $0.06~(\pm0.06)$           & $0.05~(\pm0.07)$          &$0.06~(\pm0.07)$\\
\midrule
fat tail & $\checkmark$ &  &  & $\checkmark$ & $\checkmark$ & $\checkmark$ \\
volatility clustering & $\checkmark$ &  &  & $\checkmark$ & $\checkmark$ & $\checkmark$ \\
\bottomrule
\end{tabular}
\label{Tab:result}
\end{table*}
}



\subsection{Input Data Processing}\label{Sec:input_data_processing}

Input data processing involves setting appropriate conditions that characterize the price series and transforming data into a suitable format for input to the diffusion model.

\paragraph{Calculation of Conditions}

Trend and realized volatility are considered as conditions characterizing the price series data, denoted as $p_t,~t\in\{0,...T\}$, where $p_t$ represents the price of time $t$, and $T$ denotes terminal time. By controlling the two elements, a wide range of scenarios are simulated. The trend $\hat{\mu}$ is defined as the rate of change in price. Realized volatility $\hat{\sigma}$ measures price fluctuations. Both are calculated using the logarithmic returns of price series $r_t=\log p_{t} - \log p_{t-1}, ~t\in\{1,...T\}$ as $\hat{\mu} = \sum_{t=1}^{T} r_t$, and $\hat{\sigma} = \sum_{t=1}^{T} r_{t}^2$.
In experiments, we multiply logarithmic returns by 100 before calculating each condition to rescale appropriately.


\paragraph{Wavelet Transformation}

Diffusion models, originally designed for image processing, require an additional processing step for time series. Initially, the logarithmic returns of the price series $r_t$ are standardized to $\tilde{r}_t$ with a zero mean and unit variance. Subsequently, Haar wavelet transform~\cite{wavelet} is applied to the standardized returns $\tilde{r}_t,~t\in\{1, \dots, T \}$, and the resulting coefficients are embedded in a rectangular image format. These image representations serve as inputs to the diffusion model.

\subsection{Training Phase}\label{Sec:training_phase}

During training, a conditional diffusion model is used to generate price series data under specified conditions. 

To ensure the model accurately learns the relationship between input data and conditions, cross-attention~\cite{rombach2022high} within the model is used for conditioning. Trend and realized volatility undergo affine transformations and convolutional processing to generate the key and value for cross-attention. This processing within the attention mechanism of the conditonal diffusion model enables the model to correctly learn the relationship between conditions and input data and generate corresponding data when presented with unseen conditions during inference.



\subsection{Inference Phase}\label{Sec:inference_phase}

During the inference phase, the model trained in the training phase is used to generate data aligning with specified conditions. Data generation involves two major stages. In the first stage, conditions are given as input to the conditional diffusion model for generating images based on specified conditions. 
In the second stage, the generated images are transformed back into time series data. The process is enabled by the reversibility of the Haar wavelet transform employed to convert time series data into images during training. Leveraging this property, the inverse wavelet transformation effectively reconstructs time series data from the generated images.
Thus, the process enables the generation of price series that align with any given conditions.

\section{Experimental Design}

\subsection{\takehiro{Experimental Settings}}

\paragraph{Dataset Collection}


This study used 1-minute FLEX full historical data provided by Japan Exchange Group, Inc. 

The trading hours per day was 5 \takehiro{hours}, with data spanning from January 1, 2015, to December 31, 2021. Missing data points were added using the price from the previous time step. The stocks used for the experiment are listed in Table~\ref{table:stock_codes}.

\begin{table}[h!]
\centering
\caption{List of stock codes and company names.}
\label{table:stock_codes}
\begin{tabular}{cl}
\toprule
Ticker & Company Name \\
\midrule
3407 & Asahi Kasei Corporation \\
4188 & Mitsubishi Chemical Group Corporation \\
4568 & Daiichi Sankyo Biotech Co., Ltd.\\
5020 & Eneos Corporation \\
6502 & Toshiba Corporation \\
6758 & Sony Group Corporation \\
7203 & Toyota Motor Corporation \\
7550 & Zensho Holdings Co., Ltd. \\
8306 & Mitsubishi UFJ Financial Group, Inc. \\
9202 & ANA Holdings Inc. \\
9437 & NTT Docomo, Inc. \\
\bottomrule
\end{tabular}
\end{table}


\begin{table}[t]
  \centering
  \caption{The average absolute error of synthetic data generated under Trend and Realized Volatility (RV) conditions for each generative method: CoFinDiff, vanilla GAN, and Wasserstein GAN.}
  \label{tab:gan_comparison}
  {
  \small
  \begin{tabular}{l|ccc}
    \toprule
    & vanilla GAN & Wasserstein GAN & \textbf{CoFinDiff}\\
    \midrule
    Trend & 1.83 & 1.34 & \textbf{1.02} \\
    RV    & 9.99 & 16.21 & \textbf{5.94}\\
    \bottomrule
  \end{tabular}
  }
\end{table}

\begin{figure}[t]
  \centering
  \includegraphics[width=0.9\linewidth]{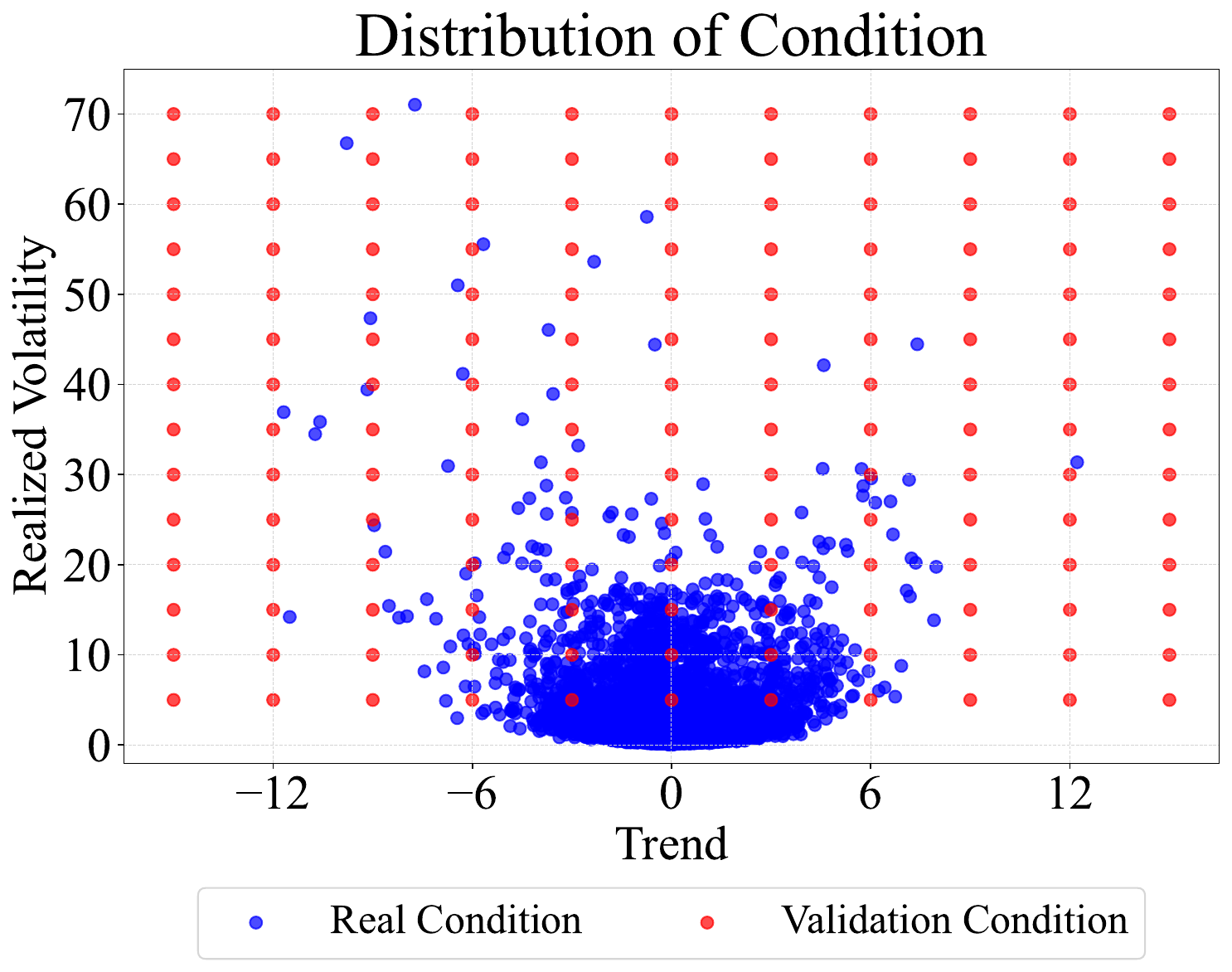}
  \caption{Scatter plots of trend and realized volatility for the conditions used in training and those used in validation.}
  \label{Fig:condition_scatter}
\end{figure}

\paragraph{Training Details}

During training, the Adam optimizer was used.
Early stopping was implemented to halt training when improvements to validation loss stopped over 100 epochs. Additionally, the number of generation steps for the diffusion model was set to 1000. The structure and parameters of the diffusion model largely followed those of the denoising diffusion probabilistic model (DDPM)~\cite{ho2020denoising}. Given the relatively small size of the input data, the number of attention heads was set to four, while the channel size of the intermediate layers was adjusted to 20.

To capture the characteristics of extreme events, upsampling was applied to the training dataset. Data corresponding to the top 25\%, 10\%, and 5\% of absolute trend values were replicated five times each. Upsampling ensured that extreme event data appeared multiple times within a single epoch. In financial markets, extreme events, although infrequent, entail substantial risk and consequently warrant attention.



\subsection{Evaluations}
\subsubsection{Baselines}


Baseline models include the GBM, GARCH model, vanilla GAN and Wasserstein GAN; in the experiment addressing RQ1, all these baselines facilitated the comparison with CoFinDiff, whereas in the experiments addressing RQ2, RQ3 and RQ4 only the GAN-based approaches (vanilla GAN and Wasserstein GAN) served as baselines.

For GBM, price path $p_t, ~t\in\{0,…,T\}$ was sampled as

\begin{equation}
    d p_t = \mu p_t dt + \sigma p_t dW_t
\end{equation}

$dt$ denotes an infinitesimal time interval, $dW_t$ denotes the corresponding increment of the Wiener process, and $\mu$ and $\sigma$ denote the drift and volatility terms respectively.
The GARCH(1, 1) model was used to generate return series. Return $r_t$ and variance $\sigma^{2}_t$ of the model were calculated, where $\epsilon_t$ adhered to a standard normal distribution.

\begin{equation}
    \begin{split}
        r_t &= \sigma_t \epsilon_t \\
        \sigma^{2}_t &= \omega + \lambda r_{t-1}^2 + \nu \sigma^{2}_{t-1}
    \end{split}
\end{equation}

where $\omega$, $\lambda$, $\nu$ are the parameters that respectively capture the baseline variance, the effect of past squared returns, and the persistence of volatility.
The structures of the vanilla GAN and Wasserstein GAN models adhere to the designs specified in ~\cite{mao2017least} and \cite{arjovsky2017wasserstein}, respectively.

\subsubsection{RQ 1: Evaluation of stylized facts}

{
\begin{table*}[t]
\centering
\caption{Diversity of synthetic data generated under specific trend and Realized Volatility (RV) conditions for each generative method: CoFinDiff, vanilla GAN, and Wasserstein GAN. We generate 200 data under the same conditions and diversity is calculated by the average of the distances in each indicator for all combinations.
($\hat{\mu}$, $\hat{\sigma}$) indicates that the given condition is trend $\hat{\mu}$ and realized volatility $\hat{\sigma}$.
}
\begin{tabular}{l|cc|cc|cc}
\toprule
& \multicolumn{2}{c|}{Wasserstein GAN} & \multicolumn{2}{c|}{vanilla GAN} & \multicolumn{2}{c}{\textbf{CoFinDiff}} \\
& (10, 50) & (-10, 50) & (10, 50) & (-10, 50) & (10, 50) & (-10, 50) \\
\midrule
DTW & $15.88(\pm5.72)$ & $14.08(\pm6.38)$ & $18.07(\pm11.56)$ & $15.91(\pm8.63)$ & $\textbf{22.58}(\pm\textbf{12.81})$ & $\textbf{16.76}(\pm\textbf{9.32})$ \\
Euclidean & $36.56(\pm15.01)$ & $25.97(\pm10.89)$ & $42.91(\pm22.33)$ & $33.35(\pm15.82)$ & $\textbf{60.30}(\pm\textbf{23.68})$ & $\textbf{52.01}(\pm\textbf{22.97})$ \\
\bottomrule
\end{tabular}
\label{Tab:diversity}
\end{table*}
}

To answer RQ 1, an experiment was performed to verify that the synthetic data generated by CoFinDiff exhibited the property of stylized facts, traditionally observed in financial time series data~\cite{stylized_facts}.
In this study, we focus on two prominent stylized facts, fat tail and volatility clustering, which are widely recognized in financial time series~\cite{volatility_clustering}.

\paragraph{Fat tail}
Fat tails, characterized by excessive probability mass in the distribution tails compared to a normal distribution, are a hallmark of financial asset returns, resulting in frequent observations of abrupt price movements. To verify the presence of fat tails in the generated data, two primary metrics were considered, Fisher's kurtosis and the Hill index. Fisher's kurtosis was calculated as

\begin{equation}
\begin{split}
    \text{Kurtosis} = \frac{T(T + 1)}{(T - 1)(T - 2)(T - 3)} \sum_{t=1}^{T} \left( \frac{r_t - \overline{r}}{s} \right)^4 \\ 
    - \frac{3(T - 1)^2}{(T - 2)(T - 3)}
\end{split}
\end{equation}

where $\overline{r}$ and $s$ denote the mean and standard deviation of time series $r_t$, respectively. Kurtosis quantified the sharpness of the distribution compared to normal distribution, with positive values indicating fat tails~\cite{kurtosis}. It was calculated for intraday returns, while the mean and standard deviation of the observed days were used for evaluation. The second measure involved the Hill index~\cite{hill}, calculated as

\small
\begin{equation}
    \xi^{\text{Hill}}_{(k(n), n)} = \left( \frac{1}{k(n)} \sum_{i=n-k(n)+1}^{n} \log \left( \frac{r_{(i,n)}}{r_{(n-k(n)+1,n)}} \right) \right) ^{-1}
\end{equation}
\normalsize
where $n$ represents the total number of observations, $k(n)$ denotes the number of extreme observations, and $r_{i, n}$ indicates the $i$th order statistic of the returns sorted in increasing order.

The Hill index served as a maximum likelihood estimator for the exponent parameter of the Pareto distribution. A typical Hill index for stock markets is known to be approximately three~\cite{cubic_law}. For this study, the Hill index was calculated using the absolute values of logarithmic returns over 1500 d, with the threshold set to top 5$\%$ of values. \hashimoto{Drawing from empirical evidences~\cite{empirical_power_law_return2,empirical_power_law_return3,empirical_power_law_return1}, we defined the stylized fact as $\xi^{\text{Hill}}_{(k(n), n)}$ in between $2.80$ and $3.40$.}

\paragraph{Volatility clustering}

Financial markets exhibit volatility clustering, a phenomenon characterized by alternating periods of high and low price volatility~\cite{volatility_clustering}. The pattern can be identified by examining the autocorrelation of absolute returns. For this study, the autocorrelation of time series data was calculated as

\begin{equation}
    \rho(\tau) = \frac{\sum_{t=1}^{T-\tau} (r_t - \overline{r})(r_{t+\tau} - \overline{r})}{\sum_{t=1}^{T} (r_t - \overline{r})^2}
\end{equation}
where $\tau$ is the time lag.

For comparative analysis, we used real data (Toyota Motor Corporation) and synthetic data obtained from the GBM, GARCH model, vanilla GAN, Wasserstein GAN, and CoFinDiff.
Results on other stocks used in the experiments are presented in Appendix \ref{stylizedfacts_realdata}.
For GBM, $\mu = 0$ and $\sigma = 1$ were set.
For GARCH model, typical parameters were set as $\omega=0.1, \lambda=0.1, \nu=0.8$ to model stock prices~\cite{5a87335d-68e7-3cf5-bb64-13813f1b0f7f}.
Conditions provided to CoFinDiff and GANs were fixed, trend = 0 and realized volatility = 1.
For consistency, the outputs of all methods were defined with 300 data points per day. The data generated by the GBM and CoFinDiff each covered 1500 day.

\subsubsection{RQ 2: Controlling generation with specific conditions}

During experimentation, the capacity of CoFinDiff to generate synthetic data adhering to various conditions was explored to answer RQ 2.
Vanilla GAN and Wasserstein GAN were selected for comparison analysis.
These models were provided with conditions not included in the training set for validation.
Figure \ref{Fig:condition_scatter} shows the distribution of conditions in both real data and data for validation.
The validation conditions were set on a grid corresponding to a wide range of trend and volatility values, aimed at assessing the ability to reproduce extreme market situations.
Evaluation of conditional generation accuracy employs mean absolute error (MAE) as the performance metric, with method comparison achieved by computing indicators corresponding to each condition from synthetic data generated under the respective conditions and subsequently calculating the MAE.

\subsubsection{RQ 3: Diversity of Synthetic Data}

An experiment was conducted to verify whether CoFinDiff generated diverse data that satisfied given conditions to answer RQ 3.
From the perspective of using synthetic data, generating diverse data that meets the necessary properties and conditions is desirable.

For the comparative analysis, we utilized both vanilla GAN and Wasserstein GAN.
We assessed the diversity of synthetic data produced by each method using dynamic time warping (DTW) and Euclidean distance.
Specifically, the conditions were fixed and 200 synthetic data samples were generated for each condition.
Further, the DTW and Euclidean distances for all possible pairs were calculated to determine the diversity of the synthetic data from their distributions.
The fixed trend and volatility conditions were set to two combinations: (10, 50) and (-10, 50)\hashimoto{, reflecting extreme events}.

\subsubsection{RQ 4: Downstream task}

Further, the effectiveness of synthetic data generated by CoFinDiff was evaluated through deep hedging tasks to answer RQ 4. 
The deep hedging task involves constructing a hedge portfolio for financial derivatives\footnote{Derivatives are financial instruments whose value is derived from the performance of an underlying asset or group of assets.} using deep learning models.
For effective hedging of financial derivatives, one must account for various factors such as transaction fees. 
Due to the extensive time required to manually take these factors into considerations, deep learning methods for hedging are gaining traction, leading to extensive research in deep hedging.
Our study aims to enhance hedge strategies by generating training data for the deep hedging task using CoFinDiff, which can produce diverse data that satisfy stylized facts and adhere to specified conditions.

The experimental setup is as follows. European call option was used as financial derivatives. A European call option is a type of financial derivative that grants the holder the right to purchase the underlying asset at a predetermined strike price\footnote{The strike price is the fixed price at which an option holder can buy or sell the underlying asset when exercising the option.} on the expiration date. The payoffs are calculated as

\begin{equation}
  \text{European Call Option Payoff} = \max(S_T - K, 0)
\end{equation}

where $S_T$ denotes the asset price at maturity $T$, and $K$ represents the strike price of the option.
The asset price processes compared include the following models: (i) CoFinDIff, (ii) vanilla GAN (iii) Wasserstein GAN (iv) real data. We select a 5-layer perceptron for learning hedge strategies, which utilizes ReLU activation function ~\cite{nair2010rectified} and layer normalization ~\cite{ba2016layer}.
The conditions for generating the synthetic data were based on the trend $\hat{\mu}$ and realized volatility $\hat{\sigma}$ of the real data. Based on these two values, we defined the conditions for generating synthetic data, and for each condition, we generate data 20 times.
The transaction cost rate was set at $1.00 \times 10^{-4}$. 
The option expiration was set to 300 minutes, consistent with previous experiments.
The prediction model takes as inputs the underlying asset price, time until maturity, the underlying asset’s position from the previous time step, volatility, and a set of indicators (delta, gamma, theta) derived from the Black-Scholes equation~\cite{black1973pricing}. For volatility, the real data adopted the previous day’s volatility, whereas the generative models utilized the realized volatility given as a condition.
For training using CoFinDiff and real data, data from January 1, 2015, to December 31, 2019, for all stocks listed in Table \ref{table:stock_codes} are utilized, and data from 2020 are used for early stop during training and data after 2021 were employed to evaluate hedge strategies.
For simplicity, all the price data used for training hedge strategies were rescaled by dividing the price at each time point by the initial price. The following two evaluation metrics were employed: (i) Entropic Risk Measure(ERM) and (ii) Conditional Value-at-Risk (CVaR).

\begin{itemize}
    \item Entropic Risk Measure:
    \begin{equation}
    \text{ERM}_{\gamma}(X) = \frac{1}{\gamma} \log \left( \mathbb{E}\left[ e^{- \gamma X} \right] \right)
    \end{equation}
    
    \item Conditional Value-at-Risk:
    \begin{equation}
    \text{CVaR}_{\alpha}(X) = \frac{1}{1-\alpha} \int_{\alpha}^{1} \text{VaR}_{u}(X) \, du
    \end{equation}
\end{itemize}

Here, $\mathbb{E}$ denotes the expected value and VaR means the risk of extreme adverse outcomes, expressed as 

\begin{equation}
    \text{VaR}_u(X) = \inf \{ x ~| ~\text{P} (X \leq x) \geq u \}
\end{equation}

where $L$ denotes the loss random variable and $u$ denotes the confidence level.

Evaluating the deep hedging task using the entire test dataset produces an average performance metric that obscures the assessment of hedging risk in individual cases. In the deep hedging task, one should examine separately those cases in which hedging plays a crucial role, namely the situations in which the absence of hedging leads to significant losses to verify the effectiveness of the hedging portfolio construction. In this context, we consider two distinct cases:
(i) The first case assumes an upward trend in the underlying asset price. In our experiment, the European call option yields a payoff only when the asset price exceeds the strike price, so a rising asset price increases risk. Accordingly, training and test data exhibiting an upward trend are used.
(ii) The second case assumes significant fluctuations in the underlying asset price. High volatility leads to large price movements and increases the likelihood of substantial payoffs, thereby raising risk as in case (i). For this case, training and test data are selected from periods with realized volatility in the top 10$\%$ or higher. To ensure neutrality in trend prediction, each synthetic dataset is normalized to a zero trend before applying volatility filtering.

\begin{table*}[t]
\centering
\caption{The results of training prediction models using four approaches — real data, CoFinDiff, vanilla GAN and Wasserstein GAN — and constructing and evaluating a hedging portfolio are presented. Here, ERM denotes the Entropic Risk Measure, and CVaR stands for Conditional Value-at-Risk. The best and second-best values are highlighted in \textbf{bold} and \textit{italic}, respectively.}
\label{tab:deephedge_result}
\begin{tabular}{ll|cccc}
\toprule
Data & Utility & Real data & vanilla GAN & Wasserstein GAN & \textbf{CoFinDiff} \\
\midrule
All data   & ERM($\gamma$=100)  & \textbf{0.00571} & 0.00624 & \textit{0.00605} & 0.00609 \\
All data   & CVaR($\alpha$=0.05)  & \textbf{0.01298} & 0.01533 & 0.01321 & \textit{0.01319} \\
Only up trend   & ERM($\gamma$=100) & \textit{-0.03149} & -0.02189 & -0.02966 & \textbf{-0.03772} \\
Only up trend   & CVaR($\alpha$=0.05)  & \textit{-0.00910} & 0.00002 & -0.00746 & \textbf{-0.01097} \\
Only high volatility    & ERM($\gamma$=100) & \textit{0.00768} & 0.00875 & 0.00790 & \textbf{0.00762} \\
Only high volatility    & CVaR($\alpha$=0.05)  & 0.01780 & 0.01830 & \textbf{0.01645} & \textit{0.01722} \\
\bottomrule
\end{tabular}
\end{table*}

\begin{figure}[t]
  \centering
  \includegraphics[width=\linewidth]{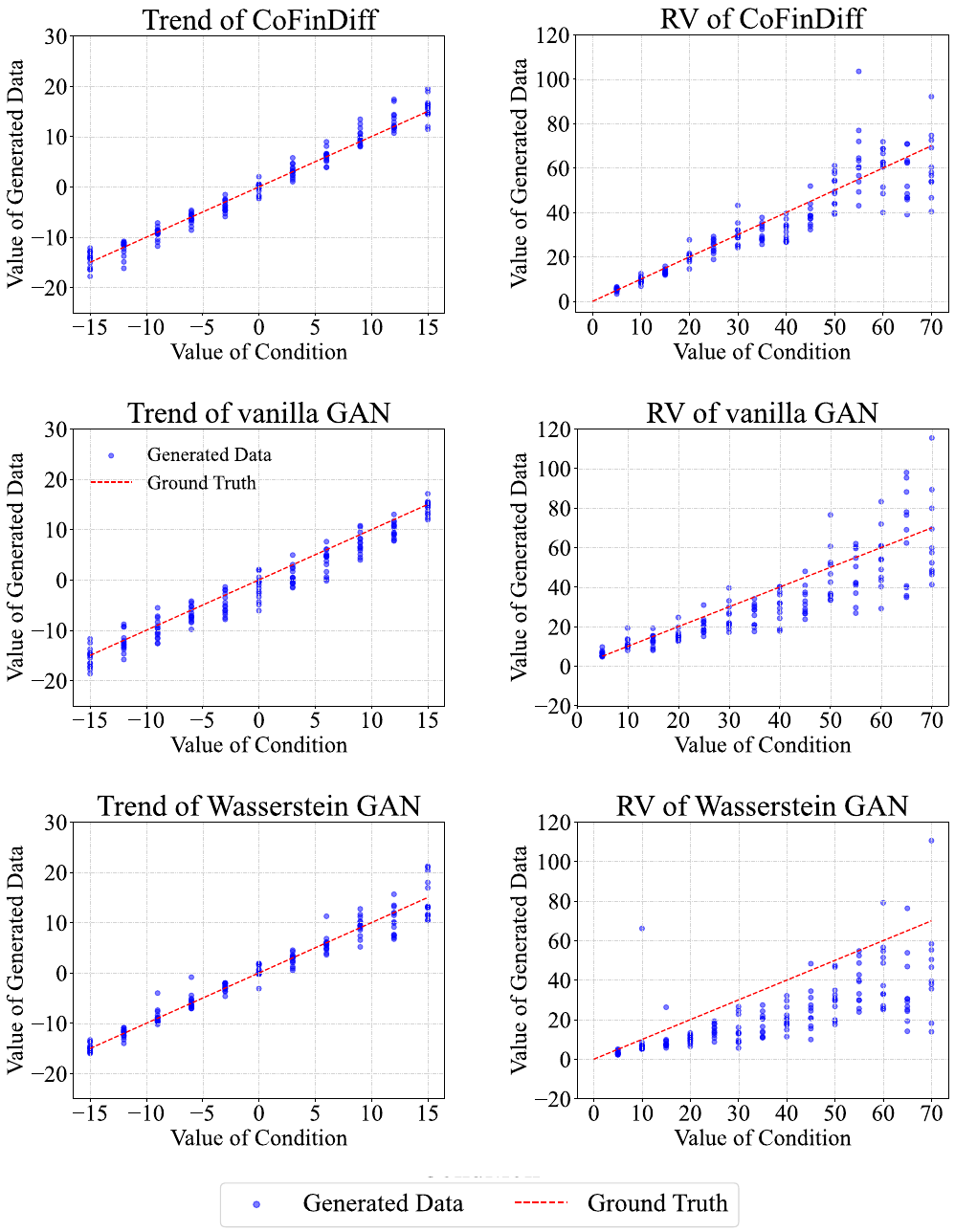}
  \caption{Scatter plots of trend and Realized Volatility (RV) conditions along with the corresponding synthetic data metrics for each generative method: CoFinDiff, vanilla GAN, and Wasserstein GAN.}
  \label{fig:scatter_plot}
\end{figure}

\section{Experimental Results and Discussion}

\subsubsection{RQ 1: Evaluation of stylized facts}

Table \ref{Tab:result} presents the calculated results for Fisher's kurtosis, Hill index, and autocorrelation in the absolute values of the return series.
The results indicate that synthetic data generated by the three models, CoFinDiff, vanilla GAN and Wasserstein GAN, satisfy the properties of two stylized facts, fat tails and volatility clustering.

\paragraph{Fat tail}

For kurtosis, the series generated by GBM and GARCH model exhibit values near zero, while the real and synthetic data generated by GANs and proposed model show positive values. Similarly, the Hill index for GBM and GARCH model exceeds the typical value of three observed in stock prices, contrary to the value of approximately three observed for real and synthetic data generated by GANs and CoFinDiff. These results indicate that the proposed conditional diffusion model successfully captured the fat tail characteristics observed in real data.

\paragraph{Volatility clustering}

While GBM exhibits negligible autocorrelation, real data and data generated by GARCH and CoFinDiff demonstrate positive autocorrelation. In particular, both real data and data generated by CoFinDiff exhibit significant positive autocorrelation, confirming the presence of volatility clustering in the latter two. Thus, the proposed conditional diffusion model successfully captured the volatility clustering phenomenon in financial markets.

\subsubsection{RQ 2: Contorolling generation with specific conditions}

Figure \ref{fig:scatter_plot} shows a scatter plot illustrating the relationship between trend and realized volatility conditions provided to each model, along with corresponding values of generated data.
The results indicate that CoFinDiff achieves higher accuracy in conditional synthetic financial data generation than vanilla GAN and Wasserstein GAN.
Although vanilla GAN exhibits notable performance under both trend conditions and realizes volatility conditions, as discussed further, the model is limited by mode collapse.
This phenomenon results in the generation of highly similar data, thereby preventing the generation of useful synthetic data.
Regarding trend conditions, both CoFinDiff and the Wasserstein GAN satisfy the conditions with similar high accuracy.
Contrarily, for realized volatility conditions, CoFinDiff generates synthetic data with higher accuracy than the Wasserstein GAN.
Specifically, when the realized volatility is high, a significant difference in accuracy between CoFinDiff and GAN is observed. 
High realized volatility indicates dramatically fluctuating prices, corresponding to extreme events. Therefore, CoFinDiff demonstrates particularly outstanding performance in generating synthetic data for extreme events, where data scarcity poses a serious challenge.

\subsubsection{RQ 3: Diversity of Synthetic Data}

Table~\ref{Tab:diversity} presents the evaluation results of diversity in synthetic data generated by each method.
The results indicate that CoFinDiff achieves the highest diversity in synthetic data. Vanilla GAN demonstrates low diversity across both evaluation metrics, which is \takehiro{likely to be} attributed to mode collapse commonly observed in GANs.
Wasserstein GAN, employing the Wasserstein distance during training to prevent mode collapse, shows some mitigation of the issue.
However, a significant disparity is observed between CoFinDiff and Wasserstein GAN under conditions corresponding to extreme events.
Therefore, CoFinDiff is capable of generating diverse data, particularly under conditions associated with extreme events.

\subsubsection{RQ 4: Downstream task}

The results are presented in Table \ref{tab:deephedge_result}. 
The results indicate that evaluations employing the complete test dataset yield superior performance for models trained on actual data, while evaluations under conditions of heightened hedging necessity reveal that models trained on data generated by CoFinDiff tend to achieve the highest performance.

When all available data served as test data, the model trained on real data achieved the best performance. One explanation attributes this result to the situation in which abundant training data remains available without assuming any specific scenario; under these circumstances, real data, which best reflects the properties of the test set, proves most suitable. In contrast, under case (i) that assumes rising underlying asset prices and case (ii) that assumes high volatility of the underlying asset, the model trained on CoFinDiff synthetic data yielded the highest performance. One explanation attributes this outcome to the reduced amount of real training data available when assuming specific scenarios; in such cases, CoFinDiff synthetic data, which provides a sufficient sample size along with high accuracy and diversity in conditions, becomes most effective.

\section{Conclusion and Future work}

This study proposes a model for generating financial time series data using a conditional diffusion model, conditioned on trends and realized volatility. The primary research question involves whether it is possible o generate diverse financial time series data that satisfy stylized facts and adhere to arbitrary conditions. To investigate this, we evaluate the following aspects: (i) whether the price series generated by the diffusion model meets stylized facts such as fat tails, (ii) whether the generated price series satisfies the specified trends and realized volatility, (iii) whether generated price series exhibit diversity, and (iv) whether in the deep hedging task, a model trained on synthetic data is capable of effectively constructing a hedging portfolio.
The experimental results confirm that the proposed conditional diffusion model meets all the criteria.

Although this study significantly contributes to the generation of financial time series data, substantial potential for further development remains. Future work can be directed toward two main areas.

First, the generation targets can be expanded. Although the current study generates data without being limited to specific stocks, it is possible to refine the model to generate data for particular stocks by conditioning on stock identifiers. Additionally, the model can be adapted to capture correlations between multiple stocks, enabling the generation of multi-stock price series. Expanding to generate other types of financial time series data, such as trading volumes or spreads, serves as a feasible direction.

The two avenues involve extending the conditions used in the generation process. This study utilizes only the daily price change rates and realized volatility as a condition. Future work can explore specifying a broader range of conditions to achieve more flexible conditional generation. Alternatively, incorporating natural language descriptions of conditions can further enhance the model's adaptability and flexibility in generating financial time series data.


\bibliographystyle{named}
\bibliography{reference}

\clearpage

\appendix
{
\noindent\Large\bfseries Appendix
}

\section{Hyperparameters}

The following lists detailed hyperparameter settings for training CoFinDiff and GANs. Training employs the Adam optimizer with an initial learning rate of $1.00 \times 10^{-4}$. The maximum number of epochs equals 3000, and a scheduler based on the cosine function decays the learning rate to $5.00 \times 10^{-6}$. Early stopping halts training after 100 consecutive epochs without a decrease in validation error; model weights corresponding to the smallest validation error receive selection. The number of generation steps for CoFinDiff equals 1000.

Detailed hyperparameter settings for the downstream task appear below. The hedge portfolio construction model employs the Adam optimizer. A uniform learning rate applies when training with real data and with synthetic data generated by CoFinDiff, vanilla GAN, or Wasserstein GAN. A learning rate of $1.00 \times 10^{-5}$ applies when using all data or only high-volatility data, whereas a learning rate of $1.00 \times 10^{-4}$ applies when using uptrend data. Early stopping activates after 100 consecutive epochs without a decrease in validation error for all data and high-volatility data, and after 1000 consecutive epochs for uptrend data. Differences in learning rate and early stopping epoch count reflect variations in convergence time and training stability.

\section{Haar wavelet transformation}

The data used in this study consists of stock price series $p_t,~t\in\{0,...T\}$. The price data is transformed into a series of log returns $r_t=\log p_{t+1} - \log p_{t}$ and then standardized. Haar wavelet transformation is applied to the standardized log returns $\tilde{r}_t,~t\in\{1,...T\}$ to obtain a series of coefficients.

In Haar wavelet transformation, we calculate the mean series $a_i$ and the difference series $d_i$ as follows. 

\begin{equation}
  \begin{split}
      a_{i+1}^{1} &= \frac{\tilde{r}_{2i} + \tilde{r}_{2i+1}}{\sqrt{2}}, \quad i = 0, 1, 2, \ldots, \bigg\lceil \frac{T}{2}-1 \bigg\rceil \\
      d_{i+1}^{1} &= \frac{\tilde{r}_{2i} - \tilde{r}_{2i+1}}{\sqrt{2}}, \quad i = 0, 1, 2, \ldots, \bigg\lceil \frac{T}{2}-1 \bigg\rceil
  \end{split}
\end{equation}

Similar operations are performed on the mean series. 

\begin{equation}
  \begin{split}
      a_{j+1}^{m+1} &= \frac{a^{m}_{2i} + a^{m}_{2i+1}}{\sqrt{2}}, \quad j = 0, 1, 2, \ldots, \bigg\lceil \frac{T}{2^{m+1}}-1 \bigg\rceil \\
      d_{j+1}^{m+1} &= \frac{a^{m}_{2i} - a^{m}_{2i+1}}{\sqrt{2}}, \quad j = 0, 1, 2, \ldots, \bigg\lceil \frac{T}{2^{m+1}}-1 \bigg\rceil
  \end{split}
\end{equation}

By repeating this process, the mean of the entire log return series and the series of difference information are ultimately obtained. These series, which have different lengths, are adjusted to match the length of the longest series $\{ d^{1}_i \}_{1}^{\big\lceil\frac{T}{2}\big\rceil}$ through appropriate repetition, converting the log return series $\bm{\tilde{r}}\in \mathbb{R}^{T}$ into 2-dimensional image $\bm{x}_0 \in \mathbb{R}^{\frac{T}{2} \times M}$, where $M = \lfloor \log_2 T \rfloor + 1$. 
The differential series obtained through a sequence of processing steps and the overall mean are embedded into a rectangular format, allowing the logarithmic return series to be treated as an image.

In our actual experiments, daily trading spanned 300 minutes, yielding a log return series \(\bm{\tilde{r}} \in \mathbb{R}^{300}\). By applying the Haar wavelet transformation to this series and performing the necessary padding, the log returns were converted into a two-dimensional image \(\bm{x}_0 \in \mathbb{R}^{152 \times 16}\). Finally, the output from the conditional diffusion model was processed by removing the padded pixels and applying the inverse wavelet transform to recover the original log return series.

\section{Diffusion model}

\begin{table*}[t]
\centering
\caption{Computed statistics representing stylized facts for individual stocks.}
\begin{tabular}{l|llllll}
\toprule
Ticker & 3407 & 4188 & 4568 & 5020 & 6502 & 6758 \\
\midrule
Kurtosis  & \(6.94~(\pm11.09)\) & \(6.33~(\pm8.54)\) & \(8.70~(\pm12.54)\) & \(5.96~(\pm8.12)\) & \(11.18~(\pm11.90)\) & \(5.96~(\pm8.13)\) \\
Hill      & \(3.11\)           & \(3.07\)           & \(2.93\)           & \(3.20\)           & \(2.83\)           & \(3.20\) \\
Acorr (1) & \(0.17~(\pm0.11)\) & \(0.20~(\pm0.10)\) & \(0.21~(\pm0.11)\) & \(0.19~(\pm0.10)\) & \(0.19~(\pm0.11)\) & \(0.19~(\pm0.10)\) \\
Acorr (5) & \(0.12~(\pm0.09)\) & \(0.14~(\pm0.09)\) & \(0.15~(\pm0.09)\) & \(0.14~(\pm0.08)\) & \(0.11~(\pm0.10)\) & \(0.14~(\pm0.08)\) \\
Acorr (10)& \(0.90~(\pm0.08)\) & \(0.10~(\pm0.08)\) & \(0.12~(\pm0.08)\) & \(0.11~(\pm0.08)\) & \(0.08~(\pm0.08)\) & \(0.11~(\pm0.08)\) \\
Acorr (20)& \(0.06~(\pm0.07)\) & \(0.71~(\pm0.07)\) & \(0.08~(\pm0.07)\) & \(0.08~(\pm0.07)\) & \(0.06~(\pm0.08)\) & \(0.08~(\pm0.07)\) \\
Acorr (30)& \(0.05~(\pm0.06)\) & \(0.06~(\pm0.07)\) & \(0.06~(\pm0.06)\) & \(0.06~(\pm0.07)\) & \(0.04~(\pm0.07)\) & \(0.06~(\pm0.07)\) \\
\bottomrule
\end{tabular}
\vspace{0.1cm}
\centering
\vspace{0.1cm}
\begin{tabular}{l|lllll}
\toprule
Ticker  & 7203 & 7550 & 8306 & 9202 & 9437 \\
\midrule
Kurtosis  & \(6.97~(\pm10.60)\) & \(6.35~(\pm6.75)\) & \(6.45~(\pm9.13)\) & \(7.97~(\pm7.97)\) & \(6.43~(\pm8.56)\) \\
Hill      & \(3.07\)           & \(3.21\)           & \(3.11\)           & \(2.95\)           & \(3.09\) \\
Acorr (1) & \(0.19~(\pm0.10)\) & \(0.14~(\pm0.09)\) & \(0.19~(\pm0.10)\) & \(0.19~(\pm0.10)\) & \(0.20~(\pm0.10)\) \\
Acorr (5) & \(0.13~(\pm0.08)\) & \(0.10~(\pm0.08)\) & \(0.14~(\pm0.09)\) & \(0.15~(\pm0.09)\) & \(0.13~(\pm0.09)\) \\
Acorr (10)& \(0.10~(\pm0.08)\) & \(0.08~(\pm0.08)\) & \(0.11~(\pm0.08)\) & \(0.11~(\pm0.08)\) & \(0.10~(\pm0.08)\) \\
Acorr (20)& \(0.07~(\pm0.07)\) & \(0.05~(\pm0.07)\) & \(0.07~(\pm0.07)\) & \(0.08~(\pm0.07)\) & \(0.06~(\pm0.07)\) \\
Acorr (30)& \(0.06~(\pm0.07)\) & \(0.04~(\pm0.06)\) & \(0.06~(\pm0.07)\) & \(0.06~(\pm0.07)\) & \(0.05~(\pm0.06)\) \\
\bottomrule
\end{tabular}
\label{Tab:result_realdata}
\end{table*}

Diffusion models learn a diffusion process that progressively adds noise to the data until it becomes pure noise, and then generate data by reversing this process.
The diffusion process, which progressively adds noise to data $x_0$ obtained by Haar wavelet transformation until it is transformed into pure noise, is expressed as follows.

\begin{equation}
  \begin{split}
    q(\bm{x}_{1:K}|\bm{x}_0) &:= \prod_{k=1}^K q(\bm{x}_k | \bm{x}_{k-1}) \\
    q(\bm{x}_k | \bm{x}_{k-1}) &:= \mathcal{N} (\sqrt{\alpha}\bm{x}_{k-1}, \beta_k \bm{I})
  \end{split}
\end{equation}

Here, the subscript $k$ represents the step number, $0<\beta_1 < \beta_2 < \cdots < \beta_K < 1$ is a hyperparameter that controls the magnitude of the variance, and $\alpha$ is a constant derived from beta that is expressed as $\alpha_k := 1 - \beta_k$.
The process of generating data by starting from noise $x_K$ and reversing the diffusion process is described as follows. 

\begin{equation}
  \begin{split}
    p_{\theta} (\bm{x}_{0:K}) &:= p_{\theta} (\bm{x}_K) \prod_{k=1}^K p_{\theta} (\bm{x}_{k-1} | \bm{x}) \\
    p_{\theta}(\bm{x}_{k-1} | \bm{x}_k) &:= \mathcal{N} (\mu_{\theta}(\bm{x}_k, k), \sigma_{k}^2) \\
    p_{\theta} (\bm{x}_K) &= \mathcal{N}(\bm{0}, \bm{I})
  \end{split}
\end{equation}

Here, $\theta$ denotes the model parameters, which are optimized using maximum likelihood estimation. The likelihood of the final sampled data $x_0$ is given by the following integral. 

\begin{equation}
  p_{\theta} (\bm{x}_0) = \int p_{\theta} (\bm{x}_{0:K}) d\bm{x}_{1:K}
\end{equation}

Due to the computationally intensive nature of this integral, optimization is performed by maximizing a variational lower bound of the log-likelihood.

\begin{equation}
\label{eq:obj1}
  \begin{split}
    - \log p_{\theta} (\bm{x}_0) &\leq \mathbb{E} \Big[ - \log \frac{p_\theta(\bm{x}_{0:K})}{q(\bm{x}_{1:K}|\bm{x}_{0})} \Big] \\
    &= \sum_{k=1} \frac{1}{2 \sigma_{t}^{2}} \mathbb{E} \Big[ || \tilde{\mu_k} (\bm{x}_k, \bm{x}_0) - \mu_{\theta} (\bm{x}_k, k) ||^2 \Big] + C
  \end{split}
\end{equation}

The objective function estimates the mean of the posterior distribution in the diffusion process by using the mean from the reverse process. Moreover, the mean in the reverse process can be expressed using the added noise as follows. 

\begin{equation}
  \mu_k(\bm{x}_k, \bm{x}_0) = \frac{1}{\sqrt{\alpha_k}} \Big( \bm{x}_k(\bm{x}_0, \epsilon) - \frac{\beta_k}{\sqrt{\bar{\beta_k}}}\epsilon \Big)
\end{equation}

Therefore, in the reverse process, instead of directly estimating the mean, the focus is on estimating the added noise using the model. 

\begin{equation}
\label{eq:mu_kheta}
  \mu_{\theta}(\bm{x}_k) = \frac{1}{\sqrt{\alpha_k}} \Big( \bm{x}_k(\bm{x}_0, \epsilon) - \frac{\beta_k}{\sqrt{\bar{\beta_k}}}\epsilon_{\theta} \Big)
\end{equation}

This approach leads to the following expression for the final objective function. 

\begin{equation}
  L(\theta) = \prod_{k=1}^K \frac{\beta_{k}^{2}}{2 \sigma_{k}^{2} \alpha_k \bar{\beta}_k} \mathbb{E} \Big[ || \epsilon - \epsilon_{\theta}(\sqrt{\bar{\alpha}_k} \bm{x}_0 + \sqrt{\bar{\beta}_k} \epsilon, k) ||^2 \Big]
\end{equation}

By learning this objective function, it becomes possible to sample by gradually removing noise through the reverse process.

\section{Conditional diffusion model}

Conditioning in diffusion models is performed using the Classifier-Free Guidance method. This technique enables conditioning without relying on classifier outputs. By weighting the outputs of the conditioned diffusion model and the unconditional diffusion model according to the following formula, data generation tailored to the conditions is achieved. 

\begin{equation}
  \nabla_{\bm{x}} \log{ p_{\gamma}(\bm{x} | y)} = \gamma \nabla_{\bm{x}} \log{p(\bm{x}| y)} + (1 - \gamma) \nabla_{\bm{x}} \log{p(\bm{x})}
\end{equation}

Here, $\bm{x}$ represents the generating data and $y$ the corresponding condition.
The parameter gamma adjusts the trade-off between adherence to the conditions and the diversity of the generated data, with higher values indicating greater fidelity to the conditions. For simplicity, $\gamma$ is set to 1 in this study.

Conditioning inputs to diffusion models are generally implemented using mechanisms such as attention. This study employs the cross-attention method, wherein conditions are transformed through linear transformations or convolutions and then used to replace the key and value in the attention mechanism.

\section{Stylized facts of Real Data}\label{stylizedfacts_realdata}

Table \ref{Tab:result_realdata} presents the indicators of the real data used for training, selected from the statistical metrics of stylized facts computed in the RQ1 evaluation but omitted from the table in the main paper.

\end{document}